\RequirePackage{fix-cm}
\documentclass[smallextended]{svjour3}       % onecolumn (second format)
\smartqed  % flush right qed marks, e.g. at end of proof

\begin{document}

\title{An improvement of a cryptanalysis algorithm%\thanks{Grants or other notes
}


\author{O. Benamara\and F. Merazka
        %Second Author %etc.
}


\institute{O. Benamara \at
               Department of mathematics  \\
              University of Science \& Technology Houari Boumediene
               P.O.Box 32 El Alia,  Algiers,  Algeria. \\
              \email{benamara.oualid@gmail.com}           %  \\
           \and
         F. Merazka \at
       Telecommunications Department\\
			University of Science \& Technology Houari Boumediene
               P.O.Box 32 El Alia,  Algiers,  Algeria. \\
              \email{fmerazka@usthb.dz}
}

\date{Received: date / Accepted: date}

\maketitle

\begin{abstract}
In this paper we present experiments in order to show how
some pseudo random number generators can improve the effectiveness
of a statistical cryptanalysis algorithm. We deduce mainly that
a better generator enhance the accuracy of the cryptanalysis algorithm.
\end{abstract}

\section{Introduction}
\label{intro}
Cryptography refers to the science that concerns encrypting data
so that, without a key, a third
person other then the sender and the receiver can not recover
the secret data \cite{Schneier}. At the same time, cryptanalist try to break
cryptosystems in order to prove that there is a security flaw. According
to Kerckhoffs' Principle, The method must not need to be kept secret, and having it fall into the enemy's hands should not cause problems \cite{Kahn}.
In this paper we deal with classical cryptosystems and try to
improve a cryptanalysis algorithm by pseudo random number generators.
Classical cryptosystems operate at the byte level of data where as the
modern one at the bit level. Pseudo random number generators have been
deeply studied for their important applications in computer science.
This paper is organized as follows. the next sections present
Markov Chain Monte Carlo (mcmc) algorithm
and a survey of the related theory. Some Pseudo Random Number Generators
(prng) with their output when
applied to mcmc are given thereafter. We analyze those results and give
our interpretation in the following section.

\section{Markov Chain Monte Carlo}
%\begin{thm}
In this section, we give the definition of Markov Chain Monte Carlo.
If a Markov chain $X_{n}$ on a finite or countable state
space $X$ is irreducible and aperiodic, with stationary
distribution $\pi$, then for every subset $A \subseteq X$,
$$\lim_{n \to \infty} P(X \in A) = \int_{A}{\pi(x) dx}$$

%\end{thm}

\subsection{MCMC algorithm }
%\begin{verbatim}
%\begin{enumerate}
In this section we rewrite the MCMC algorithm as described in  \cite{Chen}.

\begin{itemize}
\item Choose an initial state $X_{0} \in X$, where X is the
all possible states that the Marov Chain may takes.
In probability theory we call  X \textit{the universe}.
\item For n = 1, 2, 3, ...
\item Propose a new state $Y_{n} \in X$ from some symmetric
proposal density $q(X_{n-1},...,X_{0})$.
\item Let $U_{n} ~ Uniform[0, 1]$, independently of $X_{0},...,X_{n-1},Y_{n}$.
\item If $U_{n} < (\pi(Y_{n})/\pi(X_{n-1}))$, then "accept" the proposal by
setting $X_{n}=Y_{n}$, otherwise "reject" the proposal by setting $X_{n}=X_{n-1}$
\end{itemize}
%\end{enumerate}

\subsection{An MCMC algorithm to break a substitution-transposition cryptosystem}
Here is the final version of the MCMC algorithm in \cite{Chen}. The authors, after
a deep study of MCMC, chose the best parameter that output the best decryption rate:
\begin{enumerate}
\item Choose an initial state (states here are all possible encryption keys), and a fixed
scaling parameter $p>0$.
\item Repeat the following steps for many iterations (e.g. 10 000 iterations).
\begin{itemize}
\item Given the current state $x$, propose a new state $y$ from some symmetric density $q(x,y)$.
\item Sample $U_{n} ~ Uniform[0, 1]$, independently from all other variables.
\item If $u < (\pi(y)/\pi(x))^{p}$, then "accept" the proposal by replacing $x$ with $y$,
otherwise reject $y$ by leaving $x$ unchanged.
\end{itemize}
\end{enumerate}
$$\pi(x) = \prod_{\beta_{1},\beta_{2}}r(\beta_{1},\beta_{2})^{f_{x}(\beta_{1},\beta_{2})}$$
where $r$ is the frequencies of letters of the reference text and $f$ are those of
the decrypted text using the key $x$.
\subsection{Testing methodology}
We have tested the MCMC algorithm with different pseudo random number generators.
The obtained results in terms of accuracy and number of successful decryption are given in
the next paragraph.
%\end{verbatim}

\section{drand48 pseudo random number generator}
This generator generates a sequence of numbers according to this linear congruence
$$X_{n+1} =(a X_{n} + c ) mod m $$
 where $a$, $c$ and $m$ are constants.

Note that in table \ref{tab:1}, the abbreviation are as follows:
\begin{itemize}
\item EN is the experience number, which refers to the five experiences
done in this paper,
\item AC is the average correctness computed as follows. for each experience,
the text is encrypted with a different key and the cryptanalysis algorithm
is run. The cryptanalysis algorithm outputs a decryption key. The AC
measures the equality of the output of the cryptanalysis algorithm
with the actual encryption key, in the overall.
\item NSD is the number of successful decryptions, which refers to the number
of successful decryptions out of the 100 performed for each experience.
\end{itemize}

\begin{table}
\caption{drand48 prng}
\label{tab:1}       % Give a unique label
\begin{tabular}{lll}
\hline\noalign{\smallskip}
EN & AC & NSD \\
\noalign{\smallskip}\hline\noalign{\smallskip}
1 & 0.85 & 82 \\
2 & 0.85 & 82 \\
3 & 0.82 & 81 \\
4 & 0.83 & 81 \\
5 & 0.82 & 79 \\
\noalign{\smallskip}\hline
\end{tabular}
\end{table}
%\begin{tabular}{|c|c|c|}
%\hline
%experience number  & average correctness & number of successful decryption\\
%\hline
%1 & 0.85 & 82 \\
%\hline
%2 & 0.85 & 82 \\
%\hline
%3 & 0.82 & 81 \\
%\hline
%4 & 0.83 & 81 \\
%\hline
%5 & 0.82 & 79 \\
%\hline
%\end{tabular}

%We can see from \ref{tab:1} that as EN increases, AV and NSD..... This is due to .......

\section{xorshift pseudo random number generator}
One of the properties of this generator is that it is a very fast algorithm with a great period ($2^{128}-1$) \cite{Marsaglia}. Also, as we see bellow, its design is simple. Also,
it has been proved that this generator is successful in tests measuring the quality
of a pseudo random number generator.
\begin{verbatim}
#include <stdint.h>
uint32_t xor128(void) {
  static uint32_t x = 123456789;
  static uint32_t y = 362436069;
  static uint32_t z = 521288629;
  static uint32_t w = 88675123;
  uint32_t t;

  t = x ^ (x << 11);
  x = y; y = z; z = w;
  return w = w ^ (w >> 19) ^ (t ^ (t >> 8));
}
\end{verbatim}

\begin{table}
\caption{xorshift prng}
\label{tab:2}       % Give a unique label
\begin{tabular}{lll}
\hline\noalign{\smallskip}
EN & AC & NSD \\
\noalign{\smallskip}\hline\noalign{\smallskip}
1 & 0.896 & 89 \\
2 & 0.9161 & 89 \\
3 & 0.8933 & 89 \\
4 & 0.9156 & 89 \\
5 & 0.8811 & 89 \\
\noalign{\smallskip}\hline
\end{tabular}
\end{table}
%\begin{tabular}{|c|c|c|}
%\hline
%experience number  & average correctness & number of successful decryption\\
%\hline
%1 & 0.896 & 89 \\
%\hline
%2 & 0.9161 & 89 \\
%\hline
%3 & 0.8933 & 89 \\
%\hline
%4 & 0.9156 & 89 \\
%\hline
%5 & 0.8811 & 89 \\
%\hline
%\end{tabular}
Table \ref{tab:2} shows the obtained results.
%As EN increases, AV and NSD..... This is due to .......

Comparing \ref{tab:2} with \ref{tab:1} we can ay that .......

\section{chaotic iteration (CI) pseudo random number generator}
This generator is obtained from reference \cite{bgw}. As shown in this later, this generator bypass xorshift
in some tests and a deep theoretical study proved that this generator has good randomness
properties. the obtained results are tabulated in \ref{tab:3}.

Here $x$ is a binary array of length $N$.
\begin{verbatim}
a := XORshift1();
m := a mod 2 + c
for i = 0, . . . ,m do
	b := XORshift2();
	S := b mod N;
	x[S] := 1 - x[S] mod 2;
	end for
r := x;
return r;
\end{verbatim}

\begin{table}
\caption{CI prng}
\label{tab:3}       % Give a unique label
\begin{tabular}{lll}
\hline\noalign{\smallskip}
EN & AC & NSD \\
\noalign{\smallskip}\hline\noalign{\smallskip}
1 & 0.8700 & 85 \\
2 & 0.8744 & 82 \\
3 & 0.88 & 87 \\
4 & 0.8983 & 82 \\
5 & 0.8478 & 87 \\
\noalign{\smallskip}\hline
\end{tabular}
\end{table}

%\begin{tabular}{|c|c|c|}
%\hline
%experience number  & average correctness & number of successful decryption\\
%\hline
%1 & 0.8700 & 85 \\
%\hline
%2 & 0.8744 & 82 \\
%\hline
%3 & 0.88 & 87 \\
%\hline
%4 & 0.8983 & 82 \\
%\hline
%5 & 0.8478 & 87 \\
%\hline
%\end{tabular}

Comparing \ref{tab:3} with \ref{tab:2} and \ref{tab:1} we can say that
the xorshift has better results in all experiments in terms of AC and
NSD.

\section{Result discussion}
In our experiments, we have shown that with different prng, the output of the cryptanalysis
algorithm MCMC is different. More the statistical properties of the generator are better, more
the accuracy and the number of successful decryption are great. This is due to the fact that
the resulting Markov Chain has a better statistical behavior.

In the design of the MCMC algorithm, we see that prng plays a crucial role. prng
will be used in different steps of the algorithm. We use it to pick a number in $[0,1]$ uniformly and
to chose the initial state $X_{0}$. The more those parameters are random, the more Markov
Chain generated is accurate. Therefore, a good PRNG is of crucial importance in a MCMC.

Before picking any value $X_{n}$ in the Markov Chain MA, we pick a random number uniformly in
$[0,1]$ as stated in the algorithm. Naturally, a good prng will improve the convergence of
the MC and a good prng will generate a MC close to the theoretical expected result.
As the quality of a prng is measured with some standard tests, we may suppose
that a prng generating good results in our experiment is of better statistical property.
\section{Conclusion}
In this paper we presented our experiments on decryption of classical cryptosystems.
Our results show that xorshift prng are the most convenient to this kind of application since with this later we obtain the highest scores.
Previous studies showed that CI prng have the best statistical proprieties, but surprisingly
this does not implies that they are more suitable for all kind of applications like those we have done in this work.

%\begin{acknowledgements}
%\end{acknowledgements}

%\bibliography{mc.bib}   % name your BibTeX data base

\begin{thebibliography}{}
\bibitem{Schneier}
Schneier, B.: Applied Cryptography, 2nd edn.Wiley, New York (1996)

\bibitem{Kahn}
Kahn, David (second edition, 1996), The Codebreakers: the story of secret writing, Scribners p.235

\bibitem{Chen}
Chen, Jian and Rosenthal, Jeffrey S.,
Decrypting classical cipher text using Markov chain Monte Carlo ,Statistics and Computing, (2012)

\bibitem{Marsaglia}
George Marsaglia,  Xorshift RNGs,
Journal of Statistical Software, (2003-05-06)

\bibitem{bgw}
Bahi, Jacques and Guyeux, Christophe and Wang, Qianxue,  Improving random number generators by chaotic iterations. Application in data hiding,
ICCASM 2010, Int. Conf. on Computer Application and System Modeling , (2010)
\end{thebibliography}


\end{document}